\documentclass[twocolumn,aps]{revtex4}

\usepackage{graphicx} 

\begin{document}

\title{Topological cluster state quantum computing}

\author{Austin G. Fowler$^1$ and Kovid Goyal$^2$}
\affiliation{$^1$Centre for Quantum Computer Technology, University
of Melbourne, Victoria, AUSTRALIA} \affiliation{$^2$Institute for
Quantum Information, California Institute of Technology, Pasadena,
CA 91125, USA}
\date{\today}

\begin{abstract}
The quantum computing scheme described in \cite{Raus07,Raus07d},
when viewed as a cluster state computation, features a 3-D cluster
state, novel adjustable strength error correction capable of
correcting general errors through the correction of $Z$ errors only,
a threshold error rate approaching 1\% and low overhead arbitrarily
long-range logical gates.  In this work, we review the scheme in
detail framing the discussion solely in terms of the required 3-D
cluster state and its stabilizers.
\end{abstract}

\maketitle

\section{Introduction}

Classical computers manipulate bits that can be exclusively 0 or 1.
Quantum computers manipulate quantum bits (qubits) that can be
placed in arbitrary superpositions $\alpha|0\rangle +
\beta|1\rangle$ and entangled with one another $(|00\rangle +
|11\rangle)/\sqrt{2}$ \cite{Niel00}. This additional flexibility
provides both additional computing power and additional challenges
when attempting to correct the now quantum errors in the computer.
An extremely efficient scheme for quantum error correction and
fault-tolerant quantum computation is required to correct these
errors without making unphysical demands on the underlying hardware
and without introducing excessive time overhead and thus wasting a
significant amount of the potential performance increase.

This paper is a simplified review of the quantum computing scheme of
\cite{Raus07,Raus07d}.  This scheme has a number of highly desirable
properties. Firstly, it possesses a threshold error rate of 0.75\%,
meaning arbitrarily large computations can be performed arbitrarily
accurately provided the error rates of qubit initialization,
measurement, one- and two-qubit unitary gates are all less than
0.75\%. This is particularly remarkable since only nearest-neighbor
interactions between qubits are required. Furthermore, arbitrarily
distant pairs of logical qubits can be interacted with time overhead
growing only logarithmically with separation. This property enables
quantum algorithms to be implemented very efficiently \cite{vanM04}.

We have previously reviewed this scheme in 2-D \cite{Fowl08},
framing the discussion solely in terms of manipulating the
stabilizers \cite{Gott97} of the surface code \cite{Brav98}. Here we
review the scheme as a pure cluster state computation \cite{Raus03}
in 3-D, again focusing on stabilizers, without making reference to
the surface code or using the somewhat inaccessible language of
topology and homology. To further ensure broad accessibility, no
prior familiarity with cluster state quantum computation or
stabilizers is assumed.

There are many reasons to seriously consider a 3-D cluster state
approach to quantum computing. Such an approach is quite natural for
optical \cite{Knil01b,Munr05} and possibly optical lattice
\cite{Bren99,Jaks99} based quantum computing. In both cases, for
practical reasons, the 3-D cluster state would be generated and
consumed slice by slice as the computation proceeds, with just a
small number of slices, possibly just one or two, unmeasured at any
given time.  A particularly appropriate technology for generating
and measuring such a 3-D cluster state is the photonic module
\cite{Devi07}. A detailed architecture making use of the photonic
module to explicitly implement 3-D topological cluster state quantum
computing has been proposed \cite{Devi08}. An independent ion trap
architecture tailored to topological cluster states has also been
proposed \cite{Stoc08}. The existence of such architectures
underscores the need for an accessible introduction to the
underlying computation model.

The discussion is organized as follows.  In Section~\ref{Stabilizers
and cluster states}, we briefly described stabilizers and cluster
states. In Section~\ref{Topological cluster states}, we describe the
topological cluster state and give a brief overview of what
topological cluster state quantum computing involves.
Section~\ref{Logical initialization and measurement} describes
logical qubits in more detail and how to initialize them to
$|0_L\rangle$ and $|+_L\rangle$ and measure them in the $Z_L$ and
$X_L$ bases. State injection, the non-fault-tolerant construction of
arbitrary logical states, is covered in Section~\ref{State
injection}. Logical gates, namely the logical identity gate and the
logical CNOT gate, are carefully discussed in Section~\ref{Logical
gates} along with their byproduct operators.
Section~\ref{Topological cluster state error correction} describes
the error correction procedure. Section~\ref{Conclusion} concludes
and discusses some open problems.

\section{Stabilizers and cluster states}
\label{Stabilizers and cluster states}

A stabilizer \cite{Gott97} can be thought of as a convenient
notation for representing a state.  Instead of writing $|0\rangle$,
we can write $Z$ --- shorthand for the +1 eigenstate of $Z$.
Instead of $|\psi\rangle = (|00\rangle+|11\rangle)/\sqrt{2}$, we can
write $ZZ$, $XX$ --- the simultaneous eigenstate of these operators.
We will focus on states that are eigenstates of tenor products of
$I$, $X$, $Y$, $Z$.  Not all states can be described as simultaneous
eigenstates of lists of such tensor products, but a sufficiently
wide range for our purposes can be.

The result of applying a unitary gate $U$ to a state $|\psi\rangle$
is $U|\psi\rangle$.  If $|\psi\rangle$ is an eigenstate of $M$, the
new state $U|\psi\rangle$ can be written as $UMU^\dag
U|\psi\rangle$, implying $U|\psi\rangle$ is an eigenstate of
$UMU^\dag$. Given a list of stabilizers, we can thus track the
effect of gates simply by manipulating this list. Of particular
interest will be the controlled-$Z$ gate $C_Z$ which satisfies
$C_Z^\dag=C_Z$, $C_Z(I\otimes X)C_Z=Z\otimes X$, $C_Z(I\otimes
Z)C_Z=I\otimes Z$, $C_Z(X\otimes I)C_Z=X\otimes Z$ and $C_Z(Z\otimes
I)C_Z=Z\otimes I$. The action of $C_Z$ on any other stabilizer can
be determined by multiplying these relations.

A cluster state \cite{Brie01}, or more generally a graph state
\cite{Dur03}, can be defined constructively as any state obtained by
starting with a collection of qubits in $|+\rangle$ then applying
$C_Z$ gates to one or more pairs of qubits. For example, given three
qubits in $|+\rangle$, or equivalently the three stabilizers $XII$,
$IXI$, $IIX$, we can form a 1-D cluster state by applying two $C_Z$
gates to obtain the stabilizers $XZI$, $ZXZ$, $IZX$. Note that $C_Z$
gates commute. In general, a cluster state is characterized by
stabilizers of the form $X_i\otimes_{q_j\in {\rm nghb(q_i)}}Z_j$,
where $X_i$ acts on qubit $q_i$, $Z_j$ acts on qubit $q_j$ and ${\rm
nghb(q_i)}$ denotes the set of qubits connected to $q_i$ by $C_Z$
gates.

\section{Topological cluster states}
\label{Topological cluster states}

\begin{figure}
\begin{center}
\resizebox{60mm}{!}{\includegraphics{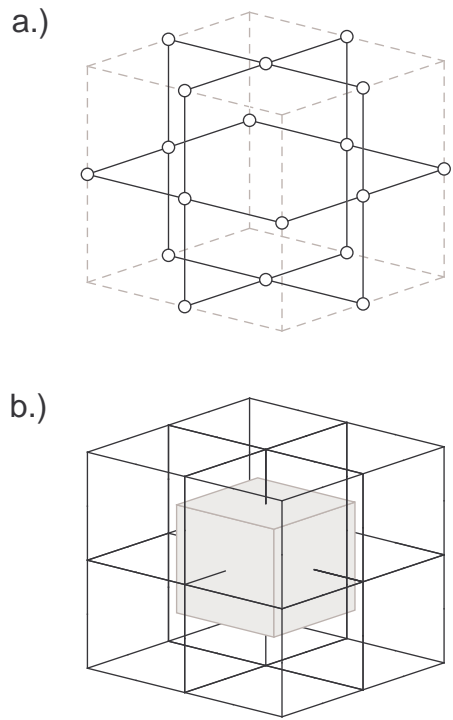}}
\end{center}
\caption{a.) A 3-D 18 qubit cluster state.  Circles denote qubits
initialized to $|+\rangle$, solid lines denote $C_Z$ gates. This
cell is tiled in 3-D to form the topological cluster state. b.) A
cube of eight cells, which we will call primal cells, each of the
form shown in part a (qubits suppressed for clarity). Location of a
dual cell (shaded) relative to its surrounding primal cells.  A dual
cell also contains exactly the arrangement of qubits shown part a.}
\label{primal_and_dual_cells}
\end{figure}

A topological cluster state is a 3-D cluster state with a specific
underlying structure. Fig.~\ref{primal_and_dual_cells}a shows a
single cell of a topological cluster state.  This cell is tiled in
3-D. Fig.~\ref{primal_and_dual_cells}b shows a simplified picture of
an 8-cell topological cluster state. The wireframe cubes as well as
the central shaded region have exactly the same form as
Fig.~\ref{primal_and_dual_cells}a. A topological cluster state can
thus be thought of as of two interlocking cubic lattices. We
arbitrarily label one of these lattices the ``primal'' lattice and
the other the ``dual'' lattice. The boundaries of the lattice are
also labeled primal or dual according to whether they consist of
primal or dual cell faces.  If we call the eight wireframe cells of
Fig.~\ref{primal_and_dual_cells}b primal cells, then the lattice has
only primal boundaries.

If we consider the product of the six stabilizers centered on the
six face qubits of Fig.~\ref{primal_and_dual_cells}a, we find that
all $Z$ operators cancel, leaving us with a cell stabilizer that is
the tensor product of $X$ on each face qubit.  This implies that if
we measure each of the face qubits in the $X$ basis, with 0
corresponding to measurement of the +1 eigenstates of $X$ and 1 the
-1 eigenstate, we will obtain six bits of information with even
parity since the individual $X$ measurements commute with the cell
stabilizer and hence the state remains a +1 eigenstate of the cell
stabilizer. A string of bits with odd parity tells us that one or
more errors have occurred locally. This is how errors are detected.
Error correction will be discussed in Section~\ref{Topological
cluster state error correction}.  For the moment we simply claim
that erroneous measurement results can be corrected arbitrarily well
given a sufficiently large topological cluster state and
sufficiently low physical error rates.  Errors are also defined to
be primal or dual according to whether they occur on primal or dual
face qubits.

\begin{figure}
\begin{center}
\resizebox{80mm}{!}{\includegraphics{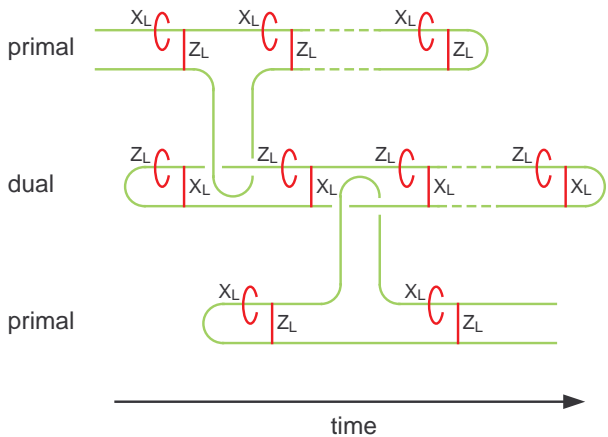}}
\end{center}
\caption{The definitions of primal and dual $Z_L$ and $X_L$ --- all
rings or chains of $Z$ operators.  Note that the logical operators
are undefined during braiding.} \label{operator_definitions}
\end{figure}

Logical qubits are associated with pairs of ``defects'' --- regions
of qubits measured in the $Z$ basis.  A defect must have a boundary
of a single type.  There are thus two types of defects and logical
qubits --- primal and dual.  Referring to
Fig.~\ref{operator_definitions}, for both primal and dual logical
qubits the initial U-shape or pair of individual beginnings
corresponds to initialization, the middle section, braided with
other defects of the opposite type, corresponds to computation and
the final U-shape or pair of individual endings corresponds to
read-out.  Full details will be given in later sections.  Logical
operators $X_L$ and $Z_L$ correspond to a ring or chain of
single-qubit $Z$ operators encircling a single defect or connecting
the pair of defects. Examples of such rings and chains are shown in
Fig.~\ref{operator_definitions}.  An important point that will
become clearer in Section~\ref{Logical initialization and
measurement} is that these rings and chains must be periodically
chosen throughout the computation --- they cannot be defined in a
consistent manner as arbitrary rings and chains since different
rings and chains are not equivalent. Furthermore, the logical
operators cannot be defined at all during braiding.  We shall always
choose primal $Z_L$ to be a chain connecting two primal defects and
dual $Z_L$ to be a ring encircling a single dual defect.  The
definitions of primal and dual $X_L$ can be inferred from
Fig.~\ref{operator_definitions}.

\begin{figure}
\begin{center}
\resizebox{70mm}{!}{\includegraphics{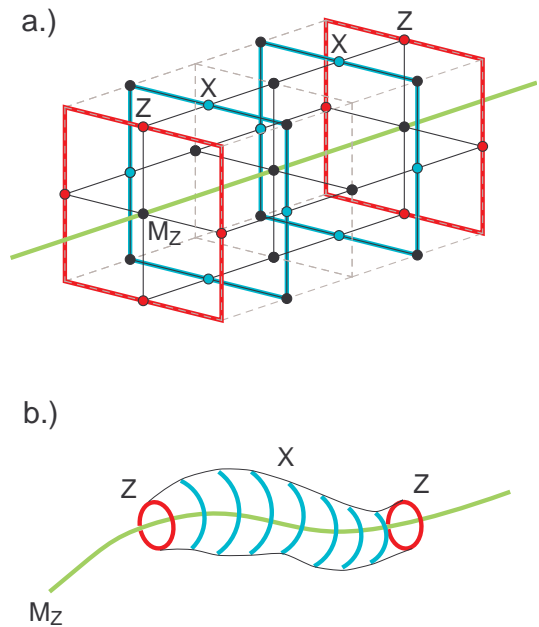}}
\end{center}
\caption{Correlation surfaces beginning and ending with rings of $Z$
operators.  a.) Red qubits are associated with $Z$ operators, blue
qubits with $X$ operators.  The collection of red and blue qubits
and their associated operators is a cluster state stabilizer.  Green
highlighting indicates qubits measured in the $Z$ basis, forming a
defect.  b.) Schematic representation.  The surface of $X$ operators
can be arbitrarily deformed whereas we keep the initial and final
rings of $Z$ operators fixed.} \label{ZZ_surface_combined}
\end{figure}

\begin{figure*}
\begin{center}
\includegraphics[width=14cm]{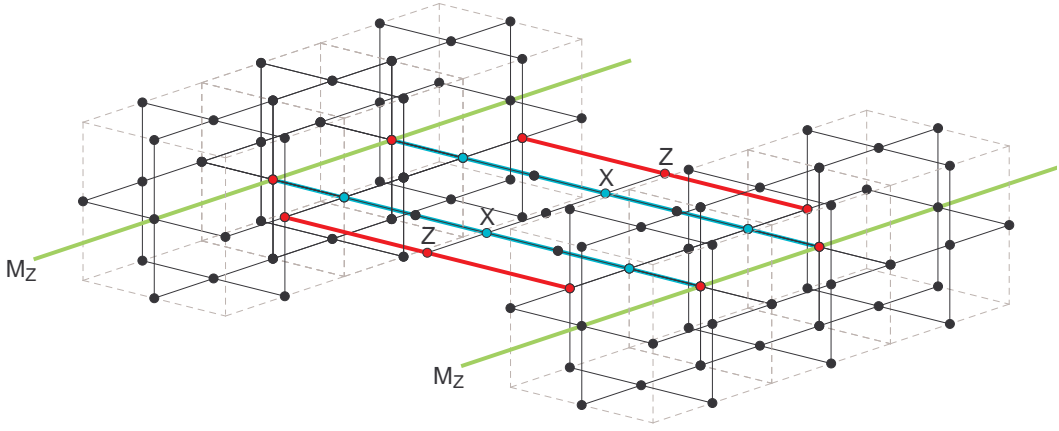}
\end{center}
\caption{A correlation surface beginning and ending with chains of
$Z$ operators.  Red qubits are associated with $Z$ operators, blue
qubits with $X$ operators.  The collection of red and blue qubits
and their associated operators is a cluster state stabilizer.  Green
highlighting indicates qubits measured in the $Z$ basis, forming
defects.} \label{XX_surface}
\end{figure*}

\begin{figure}
\begin{center}
\resizebox{50mm}{!}{\includegraphics{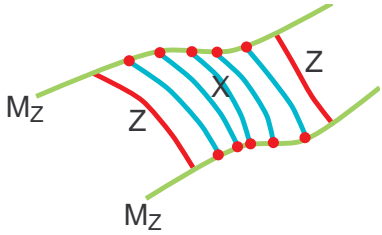}}
\end{center}
\caption{Schematic representation of Fig.~\ref{XX_surface}.  The
surface of $X$ operators can be arbitrarily deformed provided the
$Z$ operators inside the defect remain in the defect.  The initial
and final chains of $Z$ operators are kept fixed.}
\label{XX_surface_sheet}
\end{figure}

\begin{figure}
\begin{center}
\resizebox{70mm}{!}{\includegraphics{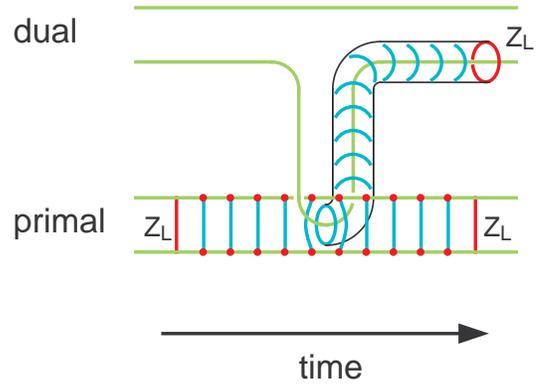}}
\end{center}
\caption{A more complicated arrangement of defects and a correlation
surface consistent with the arrangement.  The connection of $Z_L$
with $Z_LZ_L$ is suggestive of a CNOT gate, described in detail in
Section~\ref{Logical gates}.} \label{computation_example}
\end{figure}

Computation makes use of ``correlation surfaces''
--- large cluster state stabilizers connecting logical operators.
For example, two rings of $Z$ operators encircling the same defect
can be connected with a tube of $X$ operators such that a cluster
state stabilizer is formed as shown in
Fig.~\ref{ZZ_surface_combined}. Similarly, two chains of $Z$
operators connecting two defects can be connected with a surface of
$X$ operators bordered by $Z$ operators as shown in
Figs.~\ref{XX_surface}--\ref{XX_surface_sheet}.  More complicated
defect geometries lead to more complicated correlation surfaces.
Fig.~\ref{computation_example} shows two logical qubits braided in
such a way that $Z_L$ on the lower logical qubit connects to
$Z_LZ_L$ on both logical qubits --- one of the four mappings
associated with logical CNOT.  Section~\ref{Logical gates} gives
full details of the logical identity and logical CNOT gates.

\section{Logical initialization and measurement}
\label{Logical initialization and measurement}

\begin{figure}
\begin{center}
\resizebox{80mm}{!}{\includegraphics{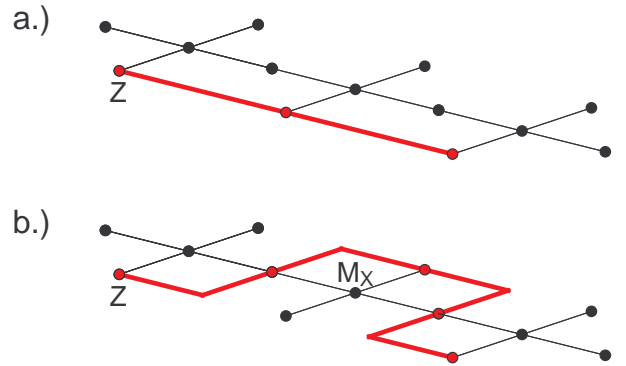}}
\end{center}
\caption{Two nonequivalent chain operators.  The second chain
operator will have an eigenvalue $(-1)^{M_X}$ times the eigenvalue
of the first chain operator.} \label{deformation}
\end{figure}

We now move on to the details of topological cluster state quantum
computing, focusing on the initialization and measurement of logical
qubits in this section.  We wish to be able to initialize logical
qubits to $|0_L\rangle$ and $|+_L\rangle$, the +1 eigenstates of
$Z_L$ and $X_L$.  Take note that deforming a logical operator does
not, in general, give an equivalent logical operator.  For example,
Fig.~\ref{deformation} shows two different chain operators. If the
lattice is in the +1 eigenstate of the first chain, it will be in
the $(-1)^{M_X}$ eigenstate of the second chain, where $M_X$ is the
result of the indicated $X$ basis measurement.  This issue can only
be avoided by periodically choosing, by hand, specific rings and
chains to represent primal and dual $Z_L$ and $X_L$.  The
correlation surfaces connecting these logical operators can,
however, take any shape consistent with the defects in the lattice.

\begin{figure*}
\begin{center}
\includegraphics[width=12cm]{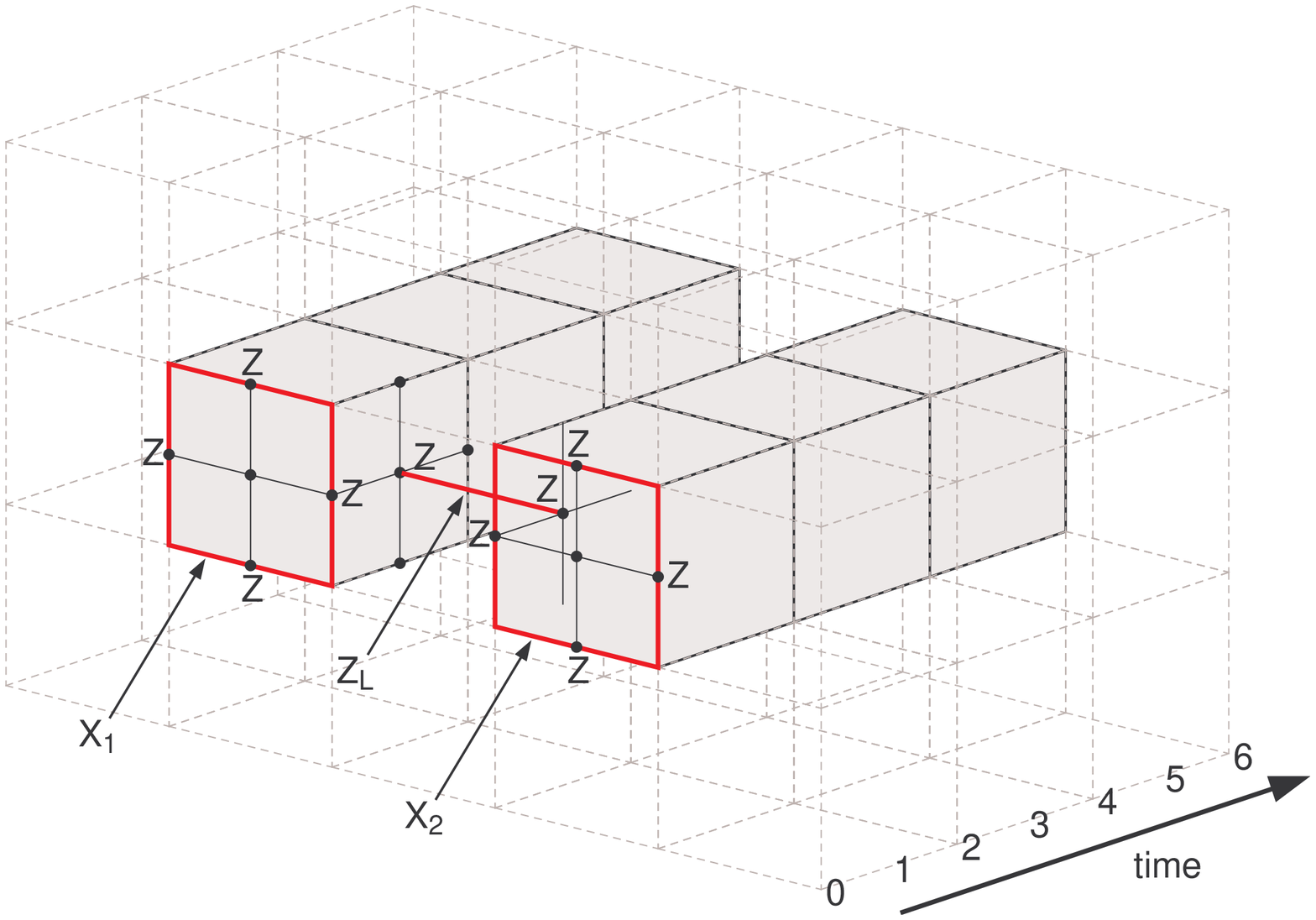}
\end{center}
\caption{A primal qubit consisting of two primal defects with
$X_L=X_1=X_2$ and $Z_L$ indicated.} \label{primal_qubit}
\end{figure*}

To permit concrete discussion, we shall choose one dimension of the
topological cluster state to be ``simulated time'' and arrange the
defects of logical qubits not currently being braided to be parallel
and in the direction of simulated time as shown in
Fig.~\ref{primal_qubit}.  Note that we define a single time step to
correspond to the measurement of a single layer of the cluster
state.  We define a primal qubit to be in the $|+_L\rangle$ state if
in a single even time step it is in the simultaneous $+1$ eigenstate
of each of the two operators consisting of a ring of single qubit
$Z$ operators encircling and on the boundary of each defect.
Similarly, the simultaneous $-1$ eigenstate of these two boundary
operators is defined to be $|-_L\rangle$.

There is some redundancy in the way we have defined $|+_L\rangle$
and $|-_L\rangle$.  It would have been sufficient to focus on a
single ring of $Z$ operators around a single defect.  Indeed,
applying both of these $Z$ rings simultaneously is the logical
identity operation
--- $X_L$ is just one of these rings, although it does not matter
which ring.  For later convenience, when we do not wish to specify
which ring, we will use the notation $X_L$. When we need to discuss
exactly which operator is being applied, we will write $X_1$ or
$X_2$.

\begin{figure*}
\begin{center}
\resizebox{150mm}{!}{\includegraphics{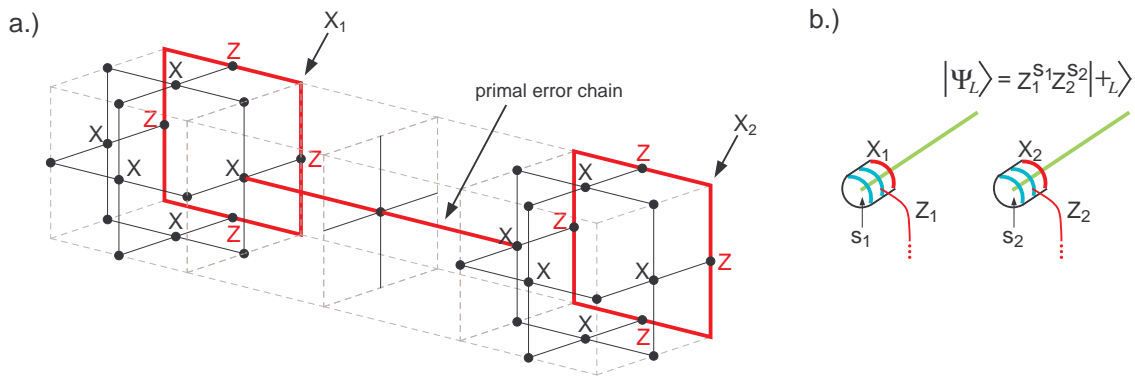}}
\end{center}
\caption{Initializing a primal qubit to the $|+_L\rangle$ state.
After the indicated $X$ measurements, the two defects are left in
known eigenstates of their associated $X_L$ operators, $X_1$ and
$X_2$, which are both rings of single-qubit $Z$ operators.}
\label{primal_+_init}
\end{figure*}

Primal qubits can be initialized to $|+_L\rangle$ up to byproduct
operators via a measurement pattern similar to that shown in
Fig.~\ref{primal_+_init}. Measuring the indicated qubits in the $X$
basis leaves the defects in either the $+1$ or $-1$ eigenstate of
$X_1$ and $X_2$ depending on the parity of the associated $X$
measurements. If we denote the parity (sum mod 2) of the $X$
measurements associated with $X_1$ by $s_1$, the state of the
logical qubit after initialization will be
$Z_1^{s_1}Z_2^{s_2}|+_L\rangle$, with $Z_L=Z_1Z_2$ and
$\{X_1,Z_1\}=\{X_2,Z_2\}=0$. The operators $Z_1$ and $Z_2$, while
not physical unless at least one additional primal boundary is
present in the system, are useful for keeping track of byproduct
operators affecting a single defect.  If an additional primal
boundary is present, these operators can be represented by chains of
$Z$ starting on each defect and ending on this additional boundary.

Note that in the absence of errors all surfaces of $X$ measurements
bounded by either $X_1$ or $X_2$ will have the same parity, as the
$X$ stabilizer associated with the six faces of a single cell can be
used to arbitrarily deform a surface without changing its parity.
This implies that the initialization procedure is well-defined and
fault-tolerant when used in conjunction with the error correction
described in Section~\ref{Topological cluster state error
correction}.

\begin{figure*}
\begin{center}
\resizebox{150mm}{!}{\includegraphics{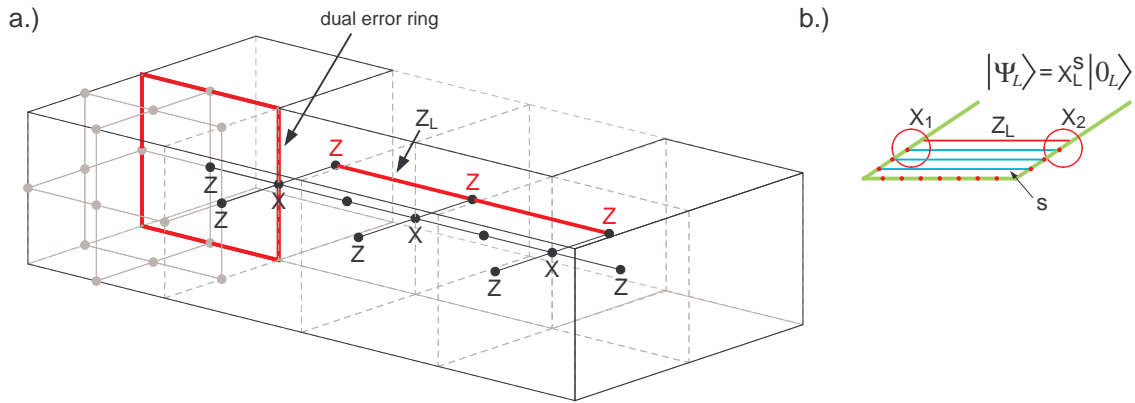}}
\end{center}
\caption{Initializing a primal qubit to the $|0_L\rangle$ state.
After the indicated $Z$ and $X$ measurements, the U-shaped defect is
left in a known eigenstate of the $Z_L$ operator, which is a
specific chosen chain of single-qubit $Z$ operators.}
\label{primal_0_init}
\end{figure*}

Primal qubits can also be initialized to $|0_L\rangle$ up to
byproduct operators via a measurement pattern similar to that shown
in Fig.~\ref{primal_0_init}.  We choose $Z_L$ to be any specific
chain of $Z$ in an odd time slice connecting two sections of defect.
The parity $s$ of the $X$ and $Z$ measurements in time slices
earlier than the chosen logical operator determines the byproduct
operator $X_L^s$.

\begin{figure*}
\begin{center}
\includegraphics[width=14cm]{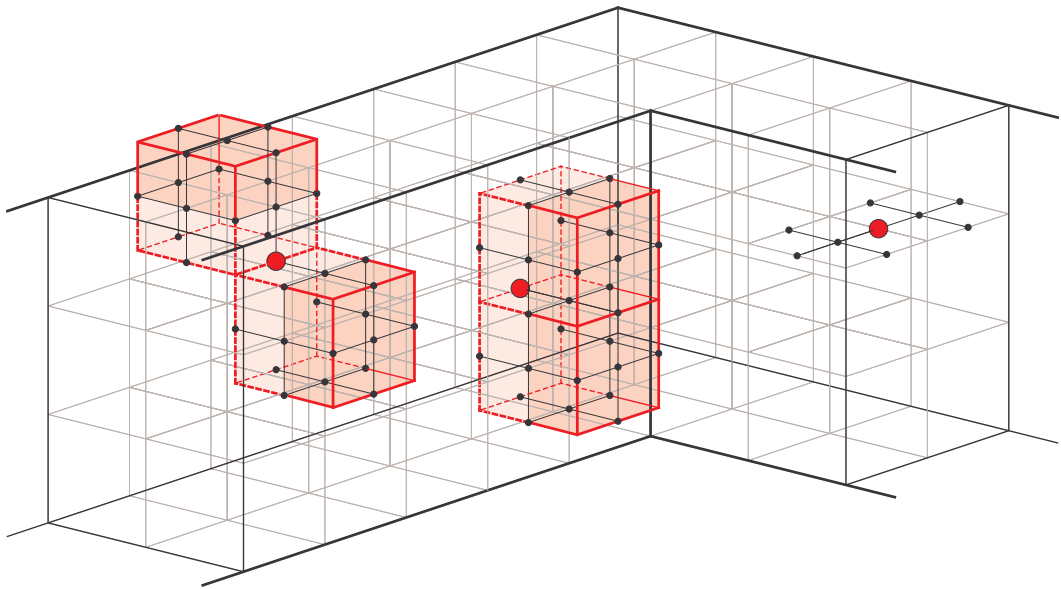}
\end{center}
\caption{Three examples of $X$ or $M_Z$ errors on qubits measured in
the $Z$ basis inside a defect.  The leftmost examples are errors on
the first layer of qubits inside the defect which can be both
detected and corrected by determining the parity of five sided dual
cells touching the error.  The rightmost example is sufficiently
deep inside the defect that no nontrivial stabilizers intersect it
and therefore the error can be ignored.}
\label{defect_surface_errors}
\end{figure*}

As drawn, Fig.~\ref{primal_0_init} is not fault-tolerant.  The
defect is too narrow to provide any information about errors on the
internal qubits measured in the $Z$ basis.
Fig.~\ref{defect_surface_errors} shows a larger defect and examples
of odd parity five sided dual cells resulting from errors on qubits
inside the defect.  For operations to be fault-tolerant, defects
must have minimum cross-section $2\times 2$ cells.

In addition to demonstrating that primal qubit initialization to
$|0_L\rangle$ can be made fault-tolerant,
Fig.~\ref{defect_surface_errors} shows how appropriate error
information is extracted on the surface of a primal defect to permit
dual error correction to continue and vice versa. Furthermore, note
that $Z$ measurements deeper inside the defect than the outermost
layer are not used in any part of the computation or error
correction procedure and as such their results can be discarded.

Dual qubit initialization, expressed in terms of dual cells, looks
absolutely identical to primal qubit initialization.  The only
difference lies in the interpretation of what the initialization
procedures mean. A dual measurement pattern of the form shown in
Fig.~\ref{primal_+_init} initializes the dual qubit to
$|0_L\rangle$. Similarly, a dual measurement pattern of the form
shown in Fig.~\ref{primal_0_init} initializes the dual qubit to
$|+_L\rangle$. The definitions of all $X$ and $Z$ logical and
byproduct operators are also reversed.

With logical qubits and logical operators defined, we can now
discuss logical errors.  In Fig.~\ref{primal_+_init}, any chain of
primal errors, which can be thought of as $Z$ errors on the
underlying qubits before measurement or $X$ basis measurement
errors, that connects the two defects is undetectable and changes
the state of the logical qubit from $|+_L\rangle$ to $|-_L\rangle$.
To make this unlikely, defects must be kept well separated.  In
Fig.~\ref{primal_0_init}, any ring of dual $Z$ and $M_X$ errors
encircling one of the defects is undetectable and changes the state
of the logical qubit from $|0_L\rangle$ to $|1_L\rangle$.  To make
this unlikely, defects must have a sufficiently large perimeter. The
situation is similar for dual qubits, with the meaning of the two
types of logical errors interchanged.

Now that we have initialization, logical measurement follows in a
straightforward manner.
Figs.~\ref{primal_+_init}--\ref{primal_0_init}, reversed in time can
be used to measure the logical operators of a qubit.  The parity of
the measurement results determines the sign of the eigenvalue of the
logical operator.

\section{State injection}
\label{State injection}

We have discussed logical qubit initialization to states
$|+_L\rangle$ and $|0_L\rangle$, measurement in the $X_L$ and $Z_L$
bases and logical errors.  We now turn our attention to state
injection, specifically the preparation of logical states
$\alpha|0_L\rangle + \beta|1_L\rangle$.

\begin{figure}
\begin{center}
\resizebox{60mm}{!}{\includegraphics{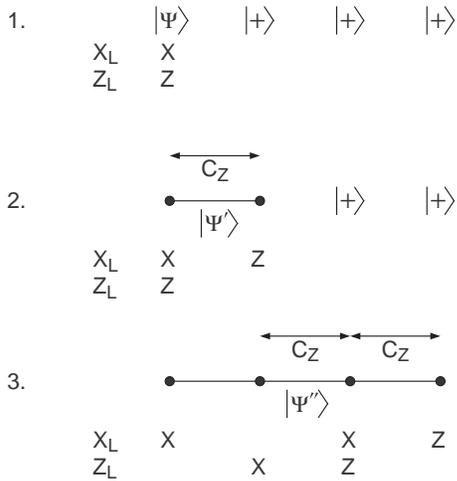}}
\end{center}
\caption{Injecting an arbitrary state into a four-qubit cluster
state. Information about the original state can be obtained from the
parity of multiple single-qubit measurements.}
\label{state_injection_simple}
\end{figure}

Consider Fig.~\ref{state_injection_simple}. The first part of the
figure shows a single qubit in an arbitrary state. The logical
operators $X_L$ and $Z_L$ correspond to single-qubit $X$ and $Z$
respectively. The second part shows the effect of applying a single
$C_Z$ gate.  A two-qubit entangled state is created, however the
parity of the two single-qubit measurements $XZ$ gives the same
information as the single-qubit measurement $X$ before the $C_Z$
gate --- $XZ$ is our new $X_L$ operator. The $C_Z$ gate transforms
+1 eigenstates of $X$ into +1 eigenstates of $XZ$. The third part of
the figure includes a further two qubits.  The essential idea is
that cluster state stabilizers centered on the second, third and
fourth qubits can be used to extend the logical operators so they
involve more qubits.

\begin{figure}
\begin{center}
\resizebox{80mm}{!}{\includegraphics{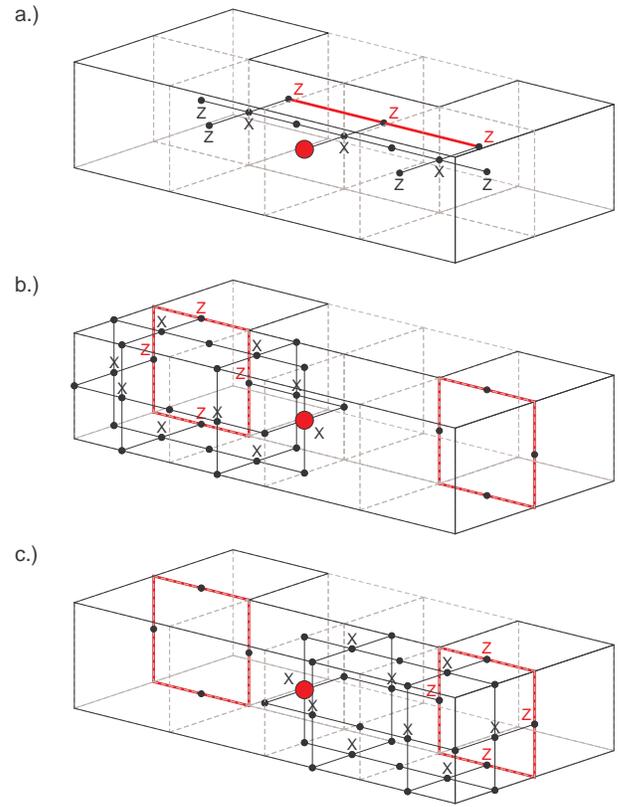}}
\end{center}
\caption{Full form of a.) $Z_L$, b.) $X_1$, c.) $X_2$ after state
injection.} \label{state_injection}
\end{figure}

Consider Fig.~\ref{state_injection}. The enlarged qubit is the
initial location of the arbitrary state.  The three parts of the
figure show $Z_L$ and $X_L=X_1=X_2$ after state injection. The
parity of the results of measuring the indicated qubits in the
indicated bases gives the same information as single-qubit
measurements on the initial state. Note that the forms of $Z_L$,
$X_1$, $X_2$ differ from those shown in Fig.~\ref{primal_qubit},
where they were simple rings and chains instead of the sheets and
socks shown in Fig.~\ref{state_injection}. This is acceptable as all
qubits associated with single-qubit operators in black are measured
in the same basis during computation implying application of these
single-qubit operators would have no effect. Nevertheless, the full
form of the logical operators is important as only from the full
form can it be seen that the logical operators anticommute.
Furthermore, measuring the single-qubit operators in black
introduces logical byproduct operators. Let $\lambda_Z$,
$\lambda_1$, $\lambda_2$ denote the parities of the measurements
indicated in black in the three parts of Fig.~\ref{state_injection}.
After measurement we will be left with the state
$X_L^{\lambda_Z}Z_1^{\lambda_1}Z_2^{\lambda_2}|\psi_L\rangle$. Note
that since state injection always begins with a single unprotected
qubit, any state injection procedure, including
Fig.~\ref{state_injection}, is inherently non-fault-tolerant.

If the enlarged qubit is prepared in an arbitrary state
$\alpha|0\rangle + \beta|1\rangle$ before being entangled with its
neighboring qubits, an arbitrary logical state $\alpha|0_L\rangle +
\beta|1_L\rangle$ can be obtained.  In practice, it is likely that
the cluster state will be prepared first, implying that the enlarged
qubit can only be rotated in the $Z$ basis as such rotations commute
with the controlled-$Z$ operators used to construct the cluster
state. Rotation before measurement could be replaced with
measurement in a rotated basis. Either way, this would limit the
class of injectable states to $(|0\rangle +
e^{i\theta}|1\rangle)/\sqrt{2}$.

\section{Logical gates}
\label{Logical gates}

Only two logical gates, the identity gate and the CNOT gate, are
required to complete the universal set of gates. These logical gates
can be understood by examining their action on logical operators.
For example, an ideal logical identity gate will have the property
$X_L\mapsto X_L$, $Z_L\mapsto Z_L$.

\begin{figure*}
\begin{center}
\resizebox{150mm}{!}{\includegraphics{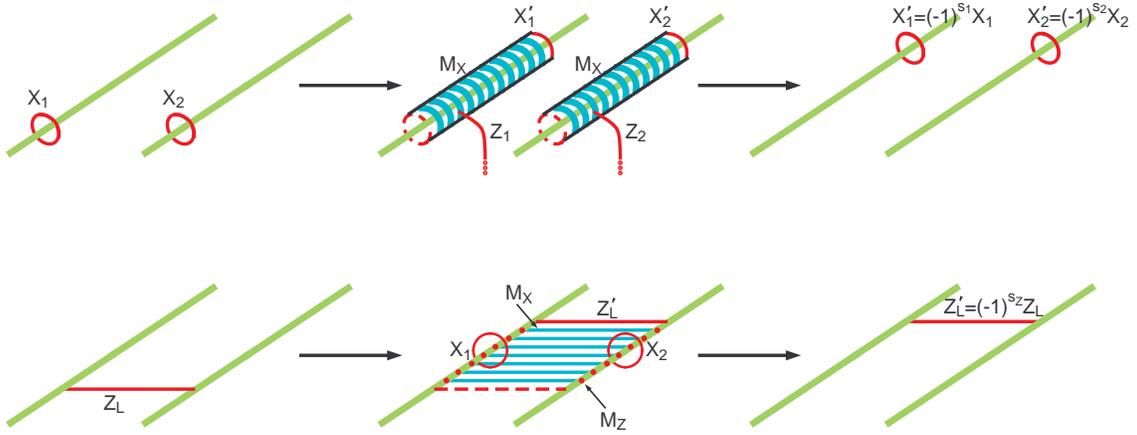}}
\end{center}
\caption{The logical identity gate. Red lines and dots indicate $Z$
operators, blue lines indicate $X$ operators. Measuring all qubits
in the middle panels in the indicated bases results in a mapping
between logical operators with byproduct operators dependent on the
parity of the measurement results.} \label{logical_identity}
\end{figure*}

Consider Fig.~\ref{logical_identity}. The upper row shows $X_1$,
$X_2$ followed by these same operators multiplied by a tubular
cluster state stabilizer. Measuring the indicated qubits in the $X$
basis results in new logical operators $X_1'=(-1)^{s_1}X_1$,
$X_2'=(-1)^{s_2}X_2$. Similarly, the lower row relates input and
output $Z_L$ via $Z_L'=(-1)^{s_Z}Z_L$. These logical operator
mappings correspond to a logical identity gate with byproduct
operators that maps logical states according to $|\psi'_L\rangle =
(I_L)Z_1^{s_1}Z_2^{s_2}X_L^{s_Z}|\psi_L\rangle$.

\begin{figure*}
\begin{center}
\resizebox{150mm}{!}{\includegraphics{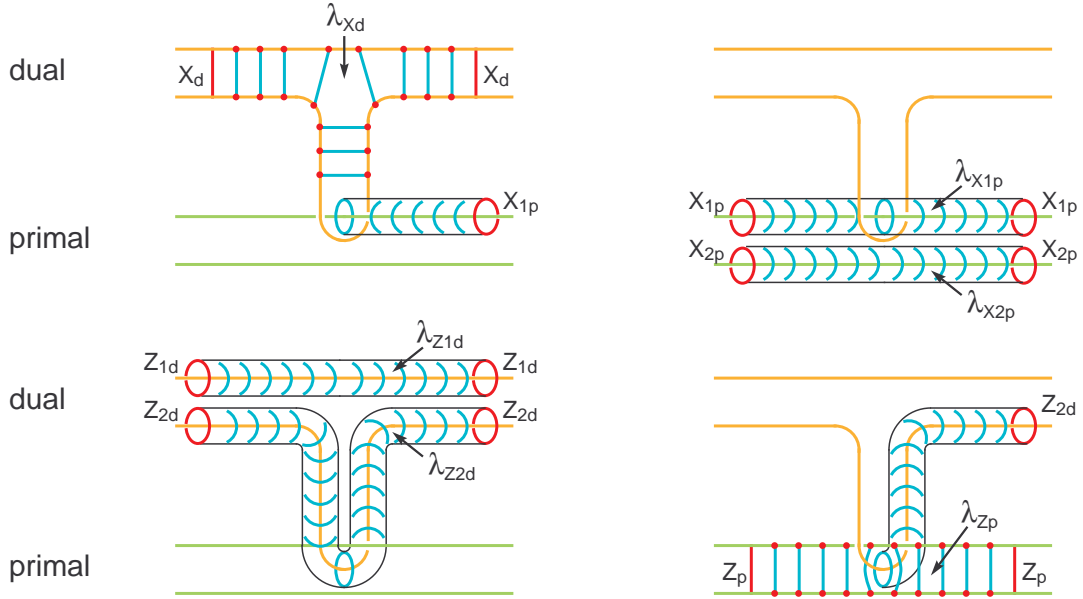}}
\end{center}
\caption{Cluster state stabilizers consistent with the indicated
braiding of defects and connecting the indicated logical operators
in a manner corresponding to logical CNOT with byproduct operators.}
\label{logical_CNOT}
\end{figure*}

Logical CNOT operates in a similar manner.  An ideal CNOT maps
control and target operators according to $X_c\mapsto X_cX_t$,
$X_t\mapsto X_t$, $Z_c\mapsto Z_c$, $Z_t\mapsto Z_cZ_t$.  Consider
Fig.~\ref{logical_CNOT}. The dual qubit is the control and the
primal qubit is the target. From the figure it can be seen that
$X_d\mapsto (-1)^{\lambda_{Xd}}X_dX_{1p}$, $X_{1p}\mapsto
(-1)^{\lambda_{X1p}}X_{1p}$, $X_{2p}\mapsto
(-1)^{\lambda_{X2p}}X_{2p}$, $Z_{1d}\mapsto
(-1)^{\lambda_{Z1d}}Z_{1d}$, $Z_{2d}\mapsto
(-1)^{\lambda_{Z2p}}Z_{2d}$, $Z_p\mapsto
(-1)^{\lambda_{Zp}}Z_{2d}Z_p$.  This corresponds to logical CNOT
with byproduct operators mapping logical states according to
\begin{equation}
|\psi'_L\rangle = C_X
Z_d^{\lambda_{Xd}}Z_{1p}^{\lambda_{X1p}}Z_{2p}^{\lambda_{X2p}}X_{1d}^{\lambda_{Z1d}}X_{2d}^{\lambda_{Z2d}}X_p^{\lambda_{Zp}}|\psi_L\rangle.
\end{equation}

\begin{figure}
\begin{center}
\resizebox{65mm}{!}{\includegraphics{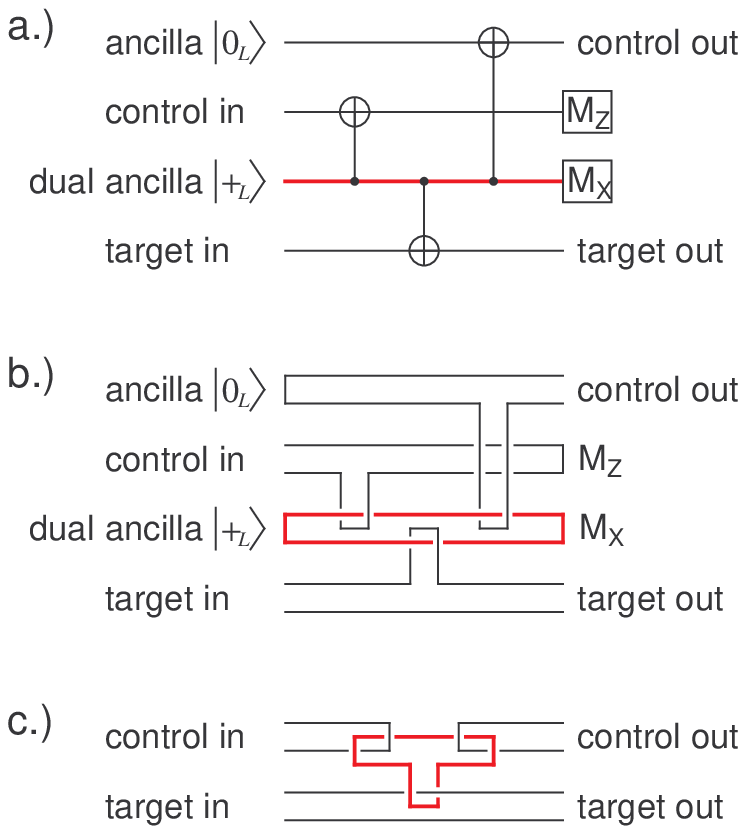}}
\end{center}
\caption{a.) Circuit comprised of logical gates described in the
text that simulates logical CNOT between two primal qubits.  b.)
Equivalent braiding of defects.  c.) Equivalent simplified braiding
of defects.} \label{CNOT_primal_primal}
\end{figure}

We do not yet quite have what we need --- a logical CNOT between two
primal qubits.  Consider Fig.~\ref{CNOT_primal_primal}a
\cite{Raus07d}.  This shows how an additional primal and dual
ancilla qubit can be used to simulate logical CNOT between two
primal qubits.  Essentially, the first CNOT and associated
measurement converts the control primal qubit into a dual qubit, the
second CNOT performs the necessary logical operation and the third
CNOT converts the dual qubit back into a primal qubit.
Fig.~\ref{CNOT_primal_primal}b shows a braiding of defects
equivalent to Fig.~\ref{CNOT_primal_primal}a and a simplified
braiding is shown in Fig.~\ref{CNOT_primal_primal}c \cite{Raus07d}.

\section{Topological cluster state error correction}
\label{Topological cluster state error correction}

Topological cluster state error correction is conceptually simple.
As discussed in Section~\ref{Topological cluster states}, measuring
the six face qubits of a given cell in the $X$ basis should yield
six bits of information with even parity. Odd parity cells indicate
the presence of errors. If we have a pair of cells with odd parity,
we can connect the cells with a path running from face qubit to face
qubit, then bit-flip the measurement results associated with the
path.  This will ensure every cell in the lattice has even parity
once more.  If we have many cells with odd parity, we can use an
efficient classical algorithm, namely the minimum weight matching
algorithm \cite{Cook99}, to pair up the cells using paths with
minimum total length. Applying bit-flips to the measurement results
along these paths again results in every cell having even parity.

\begin{figure}
\begin{center}
\resizebox{80mm}{!}{\includegraphics{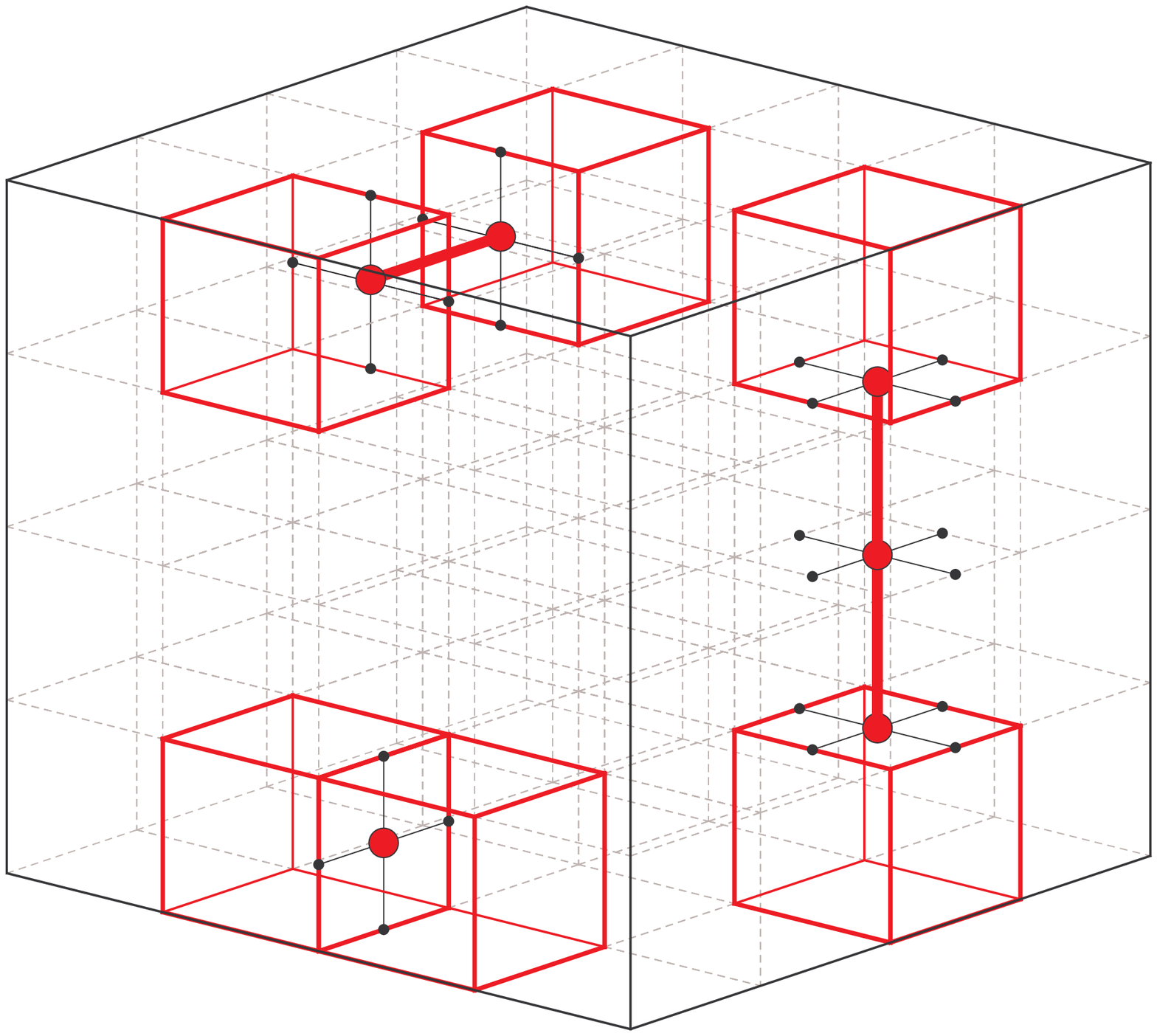}}
\end{center}
\caption{A cluster state comprised of 64 primal cells of the form
shown in Fig.~\ref{primal_and_dual_cells}a.  Three different $Z$ or
$M_X$ error chains of length 1, 2 and 3 are indicated by thick lines
and enlarged qubits.  Cells with odd parity are indicated with thick
bounding lines.  Most of the qubits in the cluster state have not
been drawn for clarity.} \label{3D_errors}
\end{figure}

There are, however, many important issues left unanswered by the
above paragraph.  Errors can occur in chains.  A lattice with 64
cells and a number of errors is shown in Fig.~\ref{3D_errors}. Only
cells at the endpoints of error chains have odd parity (indicated by
thick lines).  No information about the path of the error chain is
provided.

\begin{figure}
\begin{center}
\resizebox{65mm}{!}{\includegraphics{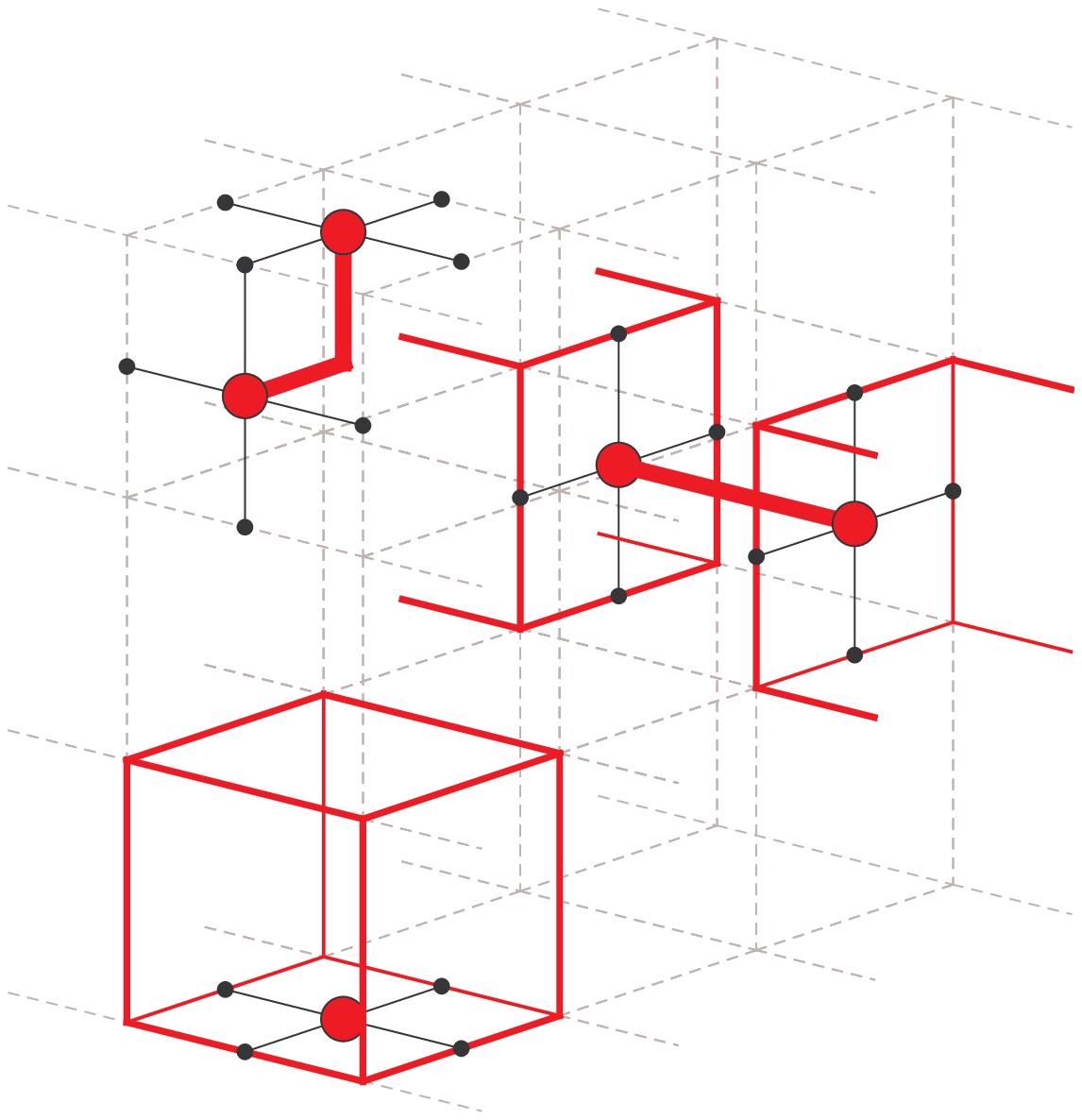}}
\end{center}
\caption{A cluster state with both primal boundaries and dual
boundaries, which consist of primal cells cut in half.  Examples of
the observable parity effects of primal error chains connected to
the two types of boundaries are included with odd parity cells
indicated by thick bounding lines.} \label{3D_errors_boundaries}
\end{figure}

The boundaries of the lattice also require special consideration.
Fig.~\ref{3D_errors} shows a primal lattice of primal cells with
primal boundaries containing primal errors.  If an endpoint of a
chain of at least two primal errors is located on a primal boundary,
the boundary cell containing this endpoint will still have even
parity.  Primal error chains that begin and end on primal boundaries
are thus undetectable and have the potential to cause logical errors
as discussed in Section~\ref{Logical initialization and
measurement}.  Fig.~\ref{3D_errors_boundaries} contains examples of
primal error chains connected to primal boundaries.
Fig.~\ref{3D_errors_boundaries} also contains dual boundaries
--- lattice boundaries that pass through the centers of primal
cells.  A primal error chain connected to a dual boundary is always
detectable as it changes the parity of the boundary cell containing
the endpoint.

\begin{figure}
\begin{center}
\resizebox{65mm}{!}{\includegraphics{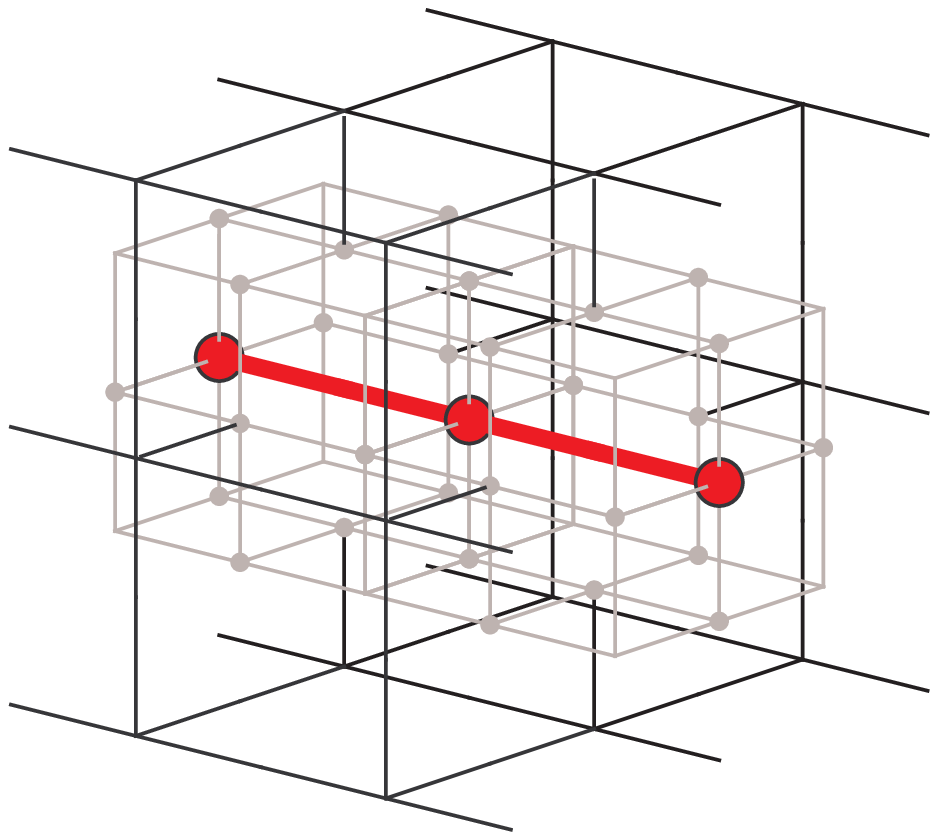}}
\end{center}
\caption{An undetectable dual error chain connecting two dual
boundaries.} \label{3D_dual_errors}
\end{figure}

Dual cells are used in an identical manner to primal cells, meaning
they also detect the presence of $Z$ or $M_X$ errors on their face
qubits. Fig.~\ref{3D_dual_errors} shows a dual error chain starting
and ending on dual boundaries.  In an analogous manner to primal
error chains, the parity of the dual boundary cells containing the
chain endpoints remains unchanged.

\begin{figure}
\begin{center}
\resizebox{50mm}{!}{\includegraphics{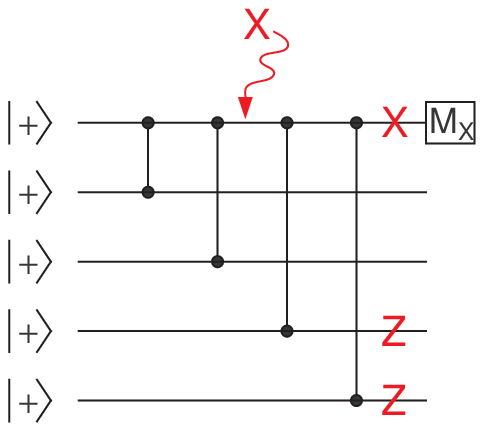}}
\end{center}
\caption{Quantum circuit showing how $X$ errors occurring at any
point during the preparation of the cluster state are equivalent to
potentially multiple $Z$ errors and an $X$ error just before
measurement in the $X$ basis, which can be ignored.}
\label{XZerrors}
\end{figure}

Primal and dual error correction occur independent of one another.
It may seem strange that both appear to only focus on $Z$ and $M_X$
errors.  An $X$ error that occurs just before an $M_X$ measurement
has no effect on the measurement result or the underlying cluster
state after the measurement.  An $X$ error that occurs during the
preparation of the cluster state, as shown in Fig.~\ref{XZerrors},
is equivalent to one or more $Z$ errors on the neighboring qubits as
well as an $X$ error just before measurement. As before, we can
ignore the $X$ error, and the error correction scheme deals with $Z$
errors.

We are now in a position to describe how correction proceeds.  Note
that only the classical measurement results will be corrected, not
any remaining unmeasured qubits.  Without loss of generality, let us
focus on primal errors.  The procedure for correcting dual errors is
analogous.  Suppose we have a connected lattice of (primal) cells
with both primal and dual boundaries.  Identify one dimension of the
lattice as simulated time.  Suppose we measure all qubits in the
lattice up to some given simulated time $t$ in the $X$ basis and
classically determine which cells in the measured region have odd
parity.  We need an algorithm to match odd parity cells with each
other and with primal boundaries such that there is a high
probability the matching corresponds to the errors that caused the
odd parity cells.  We have already mentioned that the algorithm we
will use is called the minimum weight matching algorithm
\cite{Cook99}.

The minimum weight matching algorithm takes a weighted graph with an
even number of vertices and produces a spanning list of disjoint
edges with the property that no other list has lower total weight.
The cells with odd parity become half the vertices we will feed into
the algorithm.  For every vertex in this list we add a vertex
corresponding to the nearest point on the nearest primal boundary.
We make an almost complete graph of these vertices according to the
following rules: all boundary vertices are connected to all other
boundary vertices with edge weight zero, odd parity cell vertices
are connected to all odd parity cell vertices with edge weight equal
to the sum of the absolute value of the differences of their three
coordinates measured in cells, and odd parity cell vertices are
connected to their nearest boundary vertex with edge weight equal to
the number of cells that need to be passed through to reach the
boundary plus one.  When this graph is processed by the minimum
weight matching algorithm, the resulting edge list is highly likely
to enable correction of the odd parity cells in a manner that does
not introduce logical errors.  The classical measurement results
along an arbitrary path connecting the relevant pairs of cells are
bit-flipped resulting in all measured cells having even parity. Note
that in a large computation such corrective bit-flips would only be
applied between pairs of vertices such that at least one vertex of
the pair is located at a time earlier than some $t-t_c$ where $t_c$
depends on the size of the computation.  This is to ensure that odd
parity cells close to $t$ have a chance to be matched with
appropriate partner cell, which may not yet have been measured.

\section{Conclusion}
\label{Conclusion}

We have presented a thorough review of \cite{Raus07,Raus07d},
discussing how a specific 3-D cluster state can be used to perform
general error correction despite only detecting $Z$ and $M_X$ errors
directly and detailing fault-tolerant initialization of
$|0_L\rangle$ and $|+_L\rangle$, $Z_L$ and $X_L$ measurement,
non-fault-tolerant preparation of $(|0_L\rangle +
e^{i\theta}|1_L\rangle)/\sqrt{2}$, and fault-tolerant
implementations of the logical identity gate and logical CNOT.  By
making use of state distillation \cite{Brav05,Reic05,Fowl08}, this
set of gates is sufficient to enable universal fault-tolerant
quantum computation.

Further work is required to determine the level of qubit loss the
scheme can tolerate and the dependence of the threshold error rate
on qubit loss.

\section{Acknowledgements}

We are much indebted to Robert Raussendorf for extensive and
illuminating discussions.  KG is supported by DOE Grant No.
DE-FG03-92-ER40701.

\bibliography{../../References} 

\end{document}